\documentclass[showpacs,onecolumn,12pt,aps,prb,superscriptaddress]{revtex4-1}
\usepackage{graphicx,stmaryrd,tipa,amsmath,amssymb,float,array,color,tabularx,threeparttable,booktabs,bm}
\usepackage[colorlinks,linkcolor=red,anchorcolor=red,citecolor=blue,urlcolor=red]{hyperref}
\newcommand{\etal}{\textit{et al.\/}}
\newcommand{\ie}{i.\,e.}

\begin{document}
\title{Band gap modulation in $\gamma$-graphyne by $p$-$n$ codoping}
\author{Xiaohui Deng} \email{x.deng@hynu.edu.cn}
\affiliation{Department of Physics and Electronic Information Science, Hengyang Normal University, Hengyang 421008, People's Republic of China}
\author{Jing Zeng}
\affiliation{Department of Physics and Electronic Information Science, Hengyang Normal University, Hengyang 421008, People's Republic of China}
\author{Mingsu Si}
\affiliation{Key Laboratory for Magnetism and Magnetic Materials of the Ministry of Education, Lanzhou University, Lanzhou 730000, People's Republic of China}
\author{Wei Lu}
\affiliation{University Research Facility in Materials Characterization and Device Fabrication, The Hong Kong Polytechnic University, Hong Kong, People's Republic of China}

\date{\today}
\begin{abstract}
The modulation of band gap in the two-dimensional carbon materials is of importance for their applications as electronic devices. By first-principles calculations, we propose a model to control the band gap size of $\gamma$-graphyne. The model is named as $p$-$n$ codoping, \ie, using B and N atoms to codope into $\gamma$-graphyne. After codoping, B atom plays a role of $p$ doping and N atom acts as $n$ doping. The Fermi energy level returns around the forbidden zone and the band gap of $\gamma$-graphyne vary bigger or smaller. Moreover, the gaps exhibit an oscillated behaviour in different codoping configurations. The proposed model serves as new insights for better modulation of the electronic properties of 2D carbon materials.
\end{abstract}

%81.05.Uw Carbon, diamond, graphite
%73.20.Hb Impurity and defect levels; energy states of adsorbed species
%71.15.Mb Density functional theory, local density approximation, gradient and other corrections

\maketitle
%\section{Introduction}
Two principle directions of modern semiconductors electronics are radiofrequency devices and  digital logic devices. The performance of the latter depends almost entirely on metal-oxide-semiconductor FET (MOSFET).\cite{schwierz1} For decades, silicon semiconductor has achieved great success in electronic technology. Making of smaller FET in volume is a development trend in the field of digital logic. However, its development process continuously meets the theoretical and technical challenges. There are two urgent problems in front of us, i.e thermal effect and quantum effect. Moore's law states that electronic components on the chip will increase accordingly when the central processing unit (CPU) of computer doubles in speed every 18 months.\cite{moore,schwierz2} According to this developmental rate, the quantum effects will highlight soon with the size of electronic device approximating to the nanometer size. Moreover, the reduction of size will increase the difficulty of heat dissipation of electronic devices, which may induce performance instability of devices. Therefore, looking for new alternative materials has become an important research subject of physics, chemistry and material science.

Since the successful fabrication of graphene\cite{discovery}, such systems with Dirac fermions\cite{dirac} have been under focused in both theoretical and experimental fields.\cite{graphene1,graphene2,graphene3,graphene4,graphene5} In recent years, graphyne, another kind of system with the Dirac points, is increasingly becoming attractive. Graphyne and its expanded structures are similar to graphene and can be regarded as the big hexagonal rings joined together by the acetylenic linkages (C-C$\equiv$C-C) by various construction form.\cite{Malko1,graphyne1,graphyne2,graphyne3,graphyne4,graphyne5,graphyne6,graphdiyne1,graphdiyne2} This material consists of layered carbon sheets containing {\it sp} and {\it sp}$^2$ bonds.\cite{graphyne2,graphyne6} In different construction form, the polymorph of graphyne are various including $\alpha$, $\beta$ and 6,6,12 structures and so on.\cite{Malko1,graphyne2,graphyne6,graphyne7} Most of the polymorph of graphyne are semimetal with zero gap. Malko \etal \cite{Malko1} even found that the polymorph of graphyne also possessed Dirac cones as same as graphene. As we known, In order to realize practical applications as electronic devices, a material must be introduce a band gap competitive with Si. Fortunately, a structure were prove to be direct semiconductor and its band gap is about 0.5 eV at M point in the Brillouin zone (BZ).\cite{graphyne2,graphyne3,graphyne4,graphyne6} For the convenience of narration, this structure is named as $\gamma$-graphyne. Considering the underestimation of band gap from generalized gradient approximation(GGA) calculations, Guo \etal determine the gap of $\gamma$-graphyne to be 0.96 eV from hybrid functionals.\cite{graphyne5} Therefore, from this perspective, $\gamma$-graphyne is more suitable for using in electronic devices than graphene.

Band gap modulation is a research focus correlative to graphene and graphyne. Chemical doping, molecular adsorption, substrate doping and external stress are usually the effective methods to change and control the electronic properties of carbon-related materials. Especially, the $p$-$n$ codoping method is well-established one,\cite{deng1,deng2} where an elemental doping plays a role of $p$ doping and another elemental doping acts as $n$ doping. After such codoping, the band gaps of graphene and graphyne are opened. In this work, we are aiming at the band gap modulation of $\gamma$-graphyne by $p$-$n$ codoping method. We detailed investigate the effect of dopants B and N on electronic properties and the band gap of $\gamma$-graphyne. In order to avoid the unsuccess of additional GGA calculations in determining the band gap, we reproduce the band structures of $\gamma$-graphyne with hybrid functionals. The results show that the $p$-$n$ codoping method enormously expand the band gap of $\gamma$-graphyne even in GGA level. After using the hybrid functionals, the change in band gap is more obvious. Moreover, as same as $\alpha$-graphyne, the band gaps exhibit periodic oscillations with increasing the distance of dopants.

%\section{Calculational details}
Our theoretical framework is based on the first-principles calculations. The VASP package \cite{vasp1,vasp2} is adopted to find the optimum geometries and determine their band structures of undoped,B(N)-doped and B/N-codoped graphyne. The projector-augmented wave (PAW)\cite{paw1,paw2} is used to describe the interaction between ions and electrons. GGA-type PBE functional\cite{PBE} are chosen as exchange-correlation functional. The careful tests for cutoffs of the wave function show that 500 eV is enough for wave function expansion. The $5\times5\times1$ k-points for the representation of BZ is taken to explorer the optimum geometries of considered systems. We optimize all geometries by reducing the Hellman-Feynman forces down to 0.01 ev/\AA$^2$. Based on the optimum geometries, we recalculate the energy band structures by hybrid functionals HSE06\cite{HSE06}. In both cases of band calculations from GGA-PBE and HSE06, the denser $k$-points of $11\times11\times1$ are taken for the first self-consistent step. 120 special points along high symmetry lines $\Gamma$-M-K-$\Gamma$ are chosen for band calculation and plotting. In order to address the dopant effect on the band structures of doped $\gamma$-graphyne, the electrostatic potential is first calculated and set to zero. Then the obtained eigenvalue and the Fermi energy are shifted with respect to the electrostatic potential. In the vertical direction of the sheet, a periodical boundary conditions are also set with a vacuum region of 20 {\AA} to avoid the interaction from adjacent layers. In order to quantify the stability and formation ability of the doped systems, we calculate the formation energy defined as $E_{F}=E_{DG}-E_{PG}+nE_C-\sum{E_{dopant}}$, where $E_{DG}$ is the total energy of single doped or codoped systems, $E_{PG}$ is the total energy of pristine graphyne, $E_C$ is the energy of one carbon atom in pristine graphyne, $n$ is the number of carbon atom substituted by dopants, $\sum{E_{dopant}}$ is the total energy of dopant B, N or B/N pairs.

\begin{figure*}
\begin{center}
\includegraphics[width=16cm]{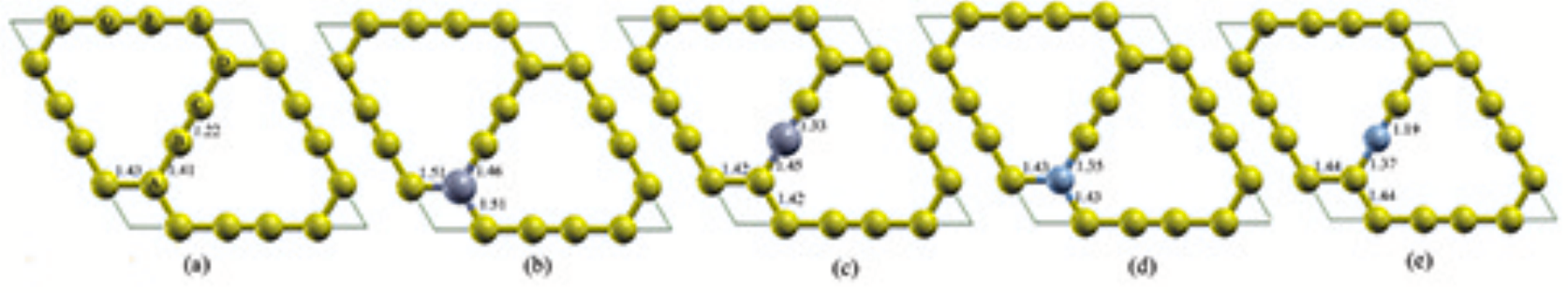}
\caption{(Color online) (a) The primitive cell of $\gamma$-graphyne. (b)-(c) The B-doped graphyne in A and B sites. (d)-(e) The N-doped graphyne in A and B sites. The numbers indicate the optimized bond lengths with the unit of \AA.  A-H indicating eight doping sites are also shown, where, A and B site are two types of inequivalent sites.}
\label{str}
\end{center}
\end{figure*}

\begin{figure}[!htbp]
\begin{center}
\includegraphics[width=16cm]{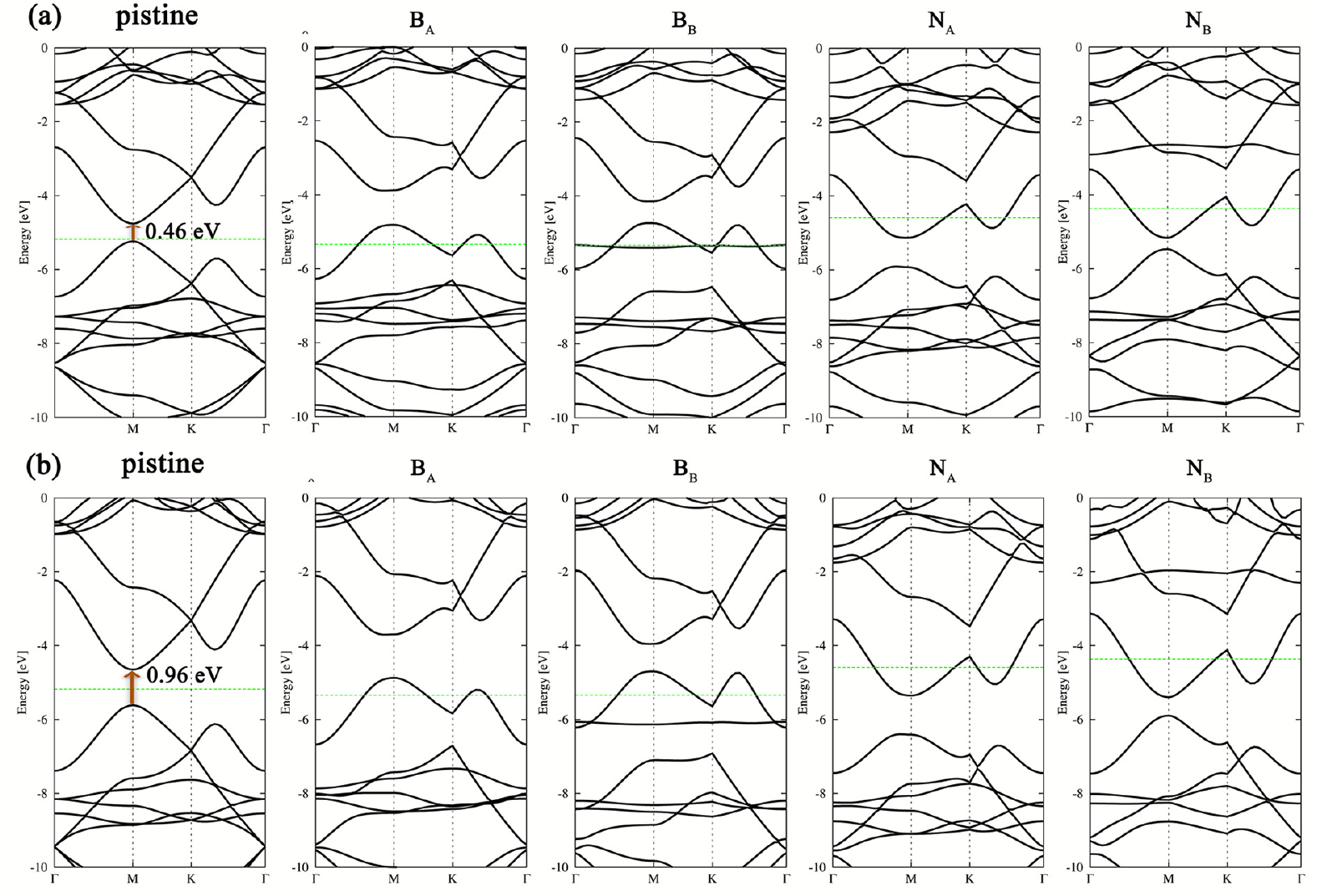}
\caption{(Color online) The band structures of pristine and B or N single codoped $\gamma$-graphyne  along $\Gamma$-M-K-$\Gamma$ line. B$_A$, B$_B$, N$_A$ and N$_B$ mean B or N atom dope in the A or B site. (a) and (b) are the PBE and HSE results, respectively. The vacuum level is set to zero and the Fermi energy level is marked by blue dotted lines.}
\label{bnd1}
\end{center}
\end{figure}

%\section{Results and discussion}
%\subsection{Single B and N doping}
The two-dimensional crystalline structure for primitive cell of $\gamma$-graphyne is hexagonal with the {\it p6/mmm} symmetry, as shown in Fig. \ref{str}(a), which is the same as graphene and other hexagonal graphyne. In $\gamma$-graphyne, the bonds among C atoms belong to C-C single bond (A-B) or C$\equiv$C triple bond (B-C). Comparing with graphene, two extra C atoms are inserted into C-C bonds of $\gamma$-graphyne, forming C-C$\equiv$C-C chain (A-B-C-D). The optimized lattice constant $a$ (or $b$) is equal to 6.89 {\AA} and the bonds of C-C and C$\equiv$C are 1.41 and 1.22 \AA, respectively, in good agreement with previous works\cite{graphyne2,graphyne4}. The calculated band structure of the pristine primitive cell is displayed in Fig.~\ref{bnd1} (a) that shows that $\gamma$-graphyne has a direct band gap of 0.46 eV at M point, being consistent with the data of 0.46 eV \cite{graphyne2} and 0.52 eV \cite{graphyne4}. Most of graphyne have Dirac corns \cite{Malko1,graphyne2,graphyne6,graphyne7}. However, $\gamma$-graphyne exhibits semiconductor behaviour with a suitable band gap, which may promise some potential applications in semiconductor devices. It is well known that the band gap of semiconductor is underestimated in the pure GGA level. However, hybrid functionals such as HSE06 can give excellent description of electronic structure of semiconductor and reproduce the experimentally measured band gap quite well. The calculation of the band gap for $\gamma$-graphyne from hybrid HSE06 functional yields a twice gap value of 0.96 eV (Fig.~\ref{bnd1} (b)). Therefore, it is necessary to adopt two functionals to confirm each other. In this paper, all the band structures for considered systems are recalculated in HSE06 level based on the optimized geometries determined by PBE functional.

The single B- and N-doped graphyne are formed by one B or N atom substituting one C atom in graphyne. There are two inequivalent sites for single B or N doping, namely, A and B sites. The structure parameters in optimized B- and N-doped graphyne are listed in Table~\ref{table1} and are shown in Fig~\ref{str}(b)-(e). For single B doping at A site, the bond lengths of three B-C are 1.46, 1.51, and 1.51 {\AA}, longer than the $d_{C-C}$ bond (1.41 {\AA}) in pristine graphyne. At B site, the long $d_{C-B}$ bond (similar to C-C bond) is 1.46 {\AA} while the short bond one is 1.33 {\AA} (similar to C$\equiv$C bond). In the case of N doping at A site, three $d_{C-N}$ are 1.35, 1.43, and 1.43 {\AA}.  At B site, the long $d_{C-N}$ bond (similar to C-C bond) is 1.37 {\AA} while the short bond one (similar to C$\equiv$C bond) is 1.19 {\AA}. As mentioned above, one can find that the C-B (C-N) bonds are longer (shorter) than the corresponding C-C or C$\equiv$C bonds both in A and B sites. It can be attributed to the fact that the N atom has more valence electrons than the B atom, leading to the shorter bond of C-N than the C-B bond. It also can be explained from the atomic radius $r$ of B, C, N with the order of $r_B>r_C>r_N$. When one B replaces one C, the B atom will repulse the adjacent C atoms outside, leading to longer C-B bond than the original C-C bond. When N replaces C, the N atom will approach the adjacent C atom due to its smaller atomic radius, leading to shorter C-N bond than the original C-C bond. The formation energy $E_{F}$ of B-doped graphyne at A and B sites are 1.78 and 2.78 eV, smaller than those of N-doped graphyne of 3.51 and 2.47 eV. In a word, B-doped $\gamma$-graphyne is more easily to synthesized in experiments than the N-doped one. To consider the doping site, the B doping prefers to select the A site while the N doping B site.

The electronic energy band structures for B- and N-doped graphyne are shown in Fig.~\ref{bnd1}, where the data for the vacuum energy level $E_{Vacuum}$, and the Fermi energy level $E_{Fermi}$ and the work function $W$ are collected in Table~\ref{table1}. For single B doping despite at A or B site, the Fermi energy level $E_Fermi$ becomes smaller comparing with pristine graphyne. Meanwhile, the valence band maximum (VBM) moves up and crosses the Fermi energy level. Therefore, B doping acts as $p$-type dopant. In the contrary, For single N doping despite at A or B site, the Fermi energy level $E_Fermi$ becomes bigger comparing with pristine graphyne. Meanwhile, the conduction band minimum (CBM) moves down and crosses the Fermi energy level. Therefore, N doping behaves as $n$-type dopant. At A site for B and N doping, the original band gap zones become obviously larger. In contrast, at B site for B and N doping, the band gap zones slightly narrows. The band structures given by PBE and HSE06 exhibit similar dispersion trend expect for the given bigger gap width from HSE06 than PBE. As the above discussions, B and N doping make the Fermi energy level move down and up relative to the forbidden band. Therefore, it will be very interesting and meaningful if the Fermi level can return to the original site using $p$-$n$ codoping, and the exiting band gap will be modulate.

%\subsection{B/N codoping}
We further investigate the geometries and electronic properties of B/N codoping in $\gamma$-graphyne to modulate its band structures. Eight possible substitutional sites, noted in Fig.~\ref{str}(a), are considered. Two types of codoping configurations are detailed investigated. They are AX and XA (X=B, C, $\cdots$, H), which represent that B (or N) atom substitutes C atom in A site and N (or B) atom substitutes C atom in B, C, D, E, F, G, and H sites, respectively.  The geometry parameters and formation energies are listed in Table~\ref{table2}. The planar structure remains after B/N codoping into graphyne. We think that the size effect plays an important role in these doped systems. B, N and C belong to the second period elements and are nearest neighbors. They have comparable atomic radius. Therefore, it is unlikely that the C atom is pulled out of the graphyne plane after B/N cooping. The $d_{C-B}$ and $d_{C-N}$ are almost the same as the results of single B and N doping. The C-B single bond covers the ranger of 1.46-1.55 {\AA} and the C$\equiv$B triple bond, similar to C$\equiv$C, is about 1.33 {\AA}. The C-N single bond covers from 1.35 to 1.42 {\AA} and the C$\equiv$B triple bond is about 1.17 {\AA}. The formation energies for the two configuration types change from 2.51 to 4.63 eV for AX and 3.56 to 5.89 eV for XA, indicating that the doped graphyne with B atom at A site is easier to synthesize than N doping at A site. For XA configuration, the GA is the most preferred codoped configuration. In two configuration types, the AB case is most easiest to form and the GA case is most difficult to realize. It means that the B-N pair induce the lowest energy in all codoped configuration owning to the charge compensation when B atom locates at A site and N atom at B site.

\begin{figure}[!htbp]
\begin{center}
\includegraphics[width=16cm]{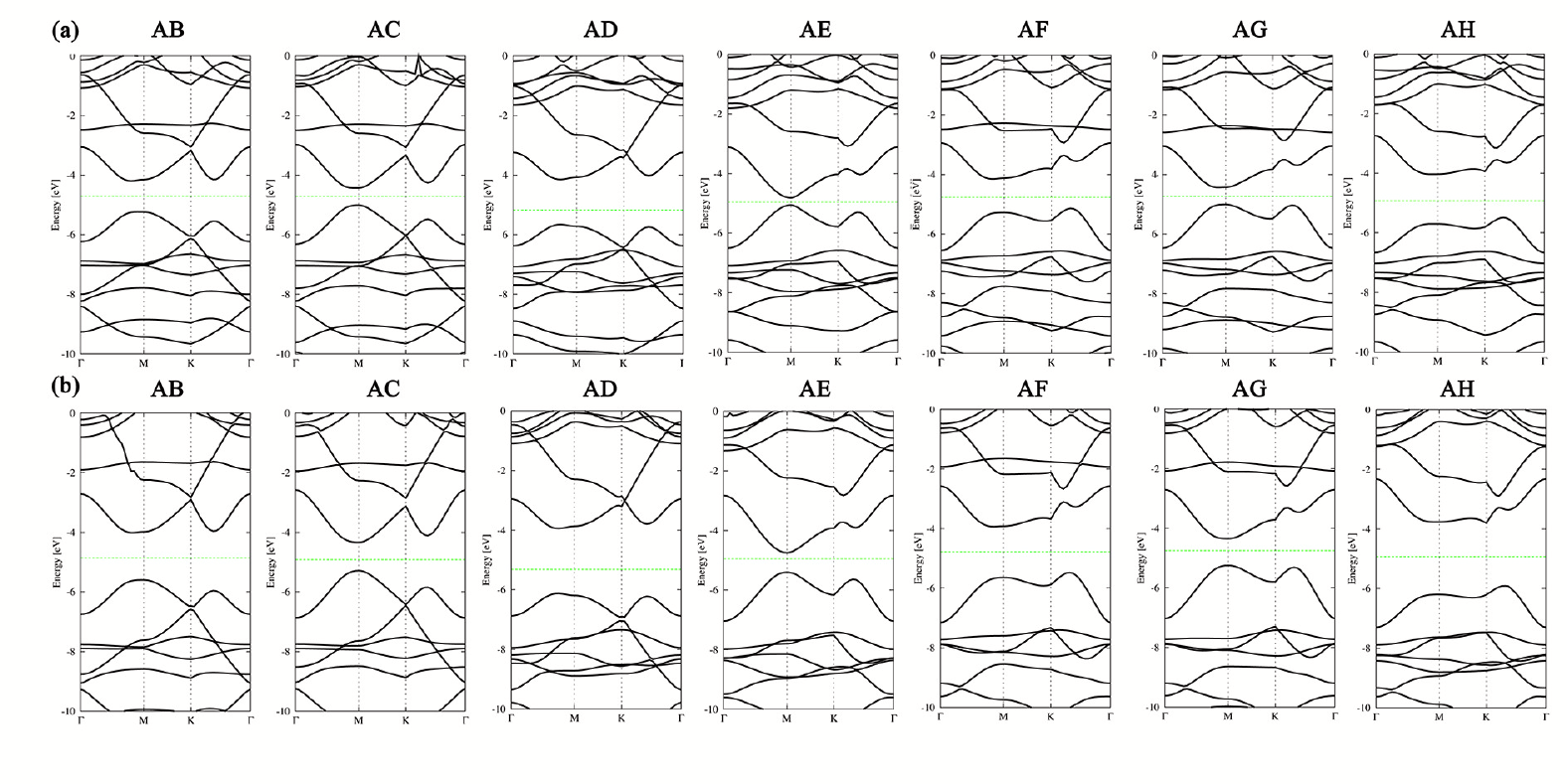}
\caption{The band structures of B/N codoped $\gamma$-graphyne for considered seven codoped configurations along $\Gamma$-M-K-$\Gamma$ line. (a) and (b) are the results in PBE and HSE level, respectively. (a) and (b) are the PBE and HSE results, respectively. The vacuum level is set to zero and the Fermi energy level is marked by blue dotted lines.}
\label{bnd2}
\end{center}
\end{figure}

\begin{figure}[!htbp]
\begin{center}
\includegraphics[width=8cm]{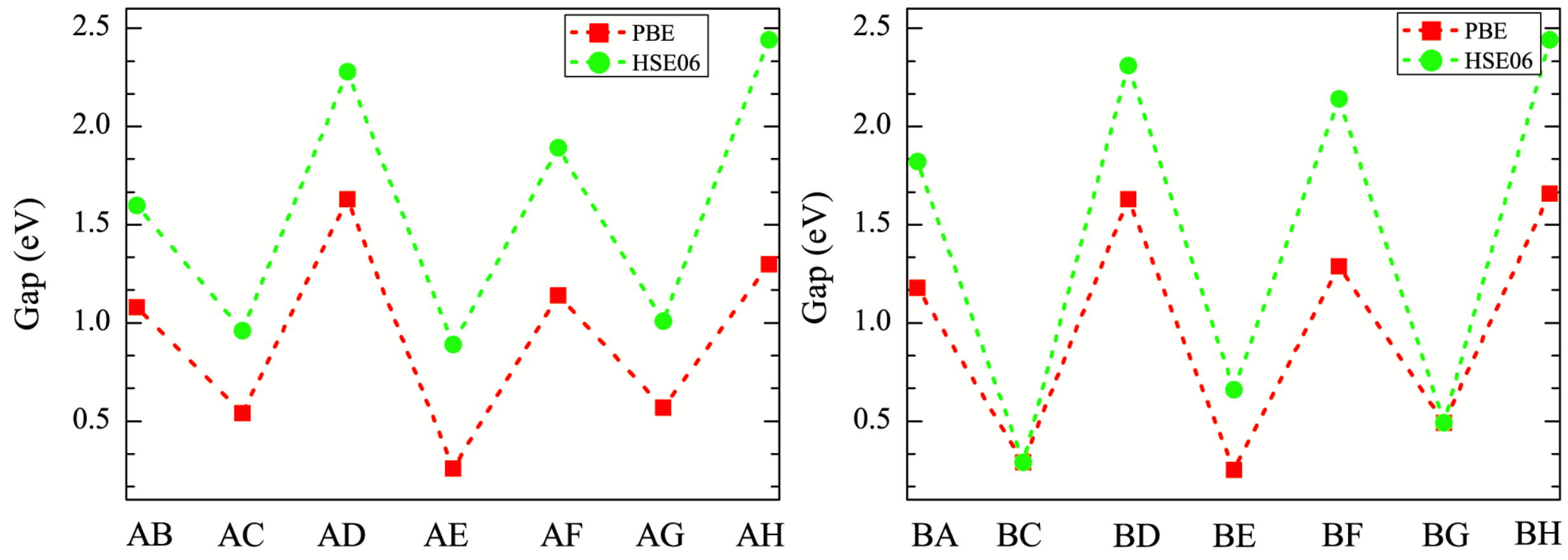}
\caption {(Color online) The oscillation of band gap for AX and XA (X=B, C, $\cdots$, H) codoping sites.}
\label{oscillation}
\end{center}
\end{figure}

The electronic band structures are calculated in PBE and HSE06 levels at last. Taking AX (X=B, C, $\cdots$, H) for example (as shown in Fig.~\ref{bnd2}), the forbidden zone of pristine $\gamma$-graphyne returns back to the Fermi energy under the counteraction between B and N doping. More importantly, around the $E_Fermi$, the different codoping site induces the different band gap after B/N codoping. Furthmore, the band gaps exhibit the oscillation behavior, changing in the big-small-big way. The situation of XA (X=B, C, $\cdots$, H) are similar to XA with the oscillated band gap. The periodic oscillation behavior exited in B/N codoped graphyneas, shown in Fig.~\ref{oscillation}, may be due to collaborative effect between the the symmetry breaking and energy level coupling. Different codoping configurations will break the symmetry of $\gamma-$graphyne in different way, and thus introduce variable band gaps. On another hand, the $p-n$ codoping from B and N atoms impose two opposite influence on the Fermi energy of undoped $\gamma-$graphyne. This coupling each other also bring different band gap size to system. The band gaps values for AX and XA are summarized in Table~\ref{table2}. For AX codoping configuration, the induced gaps in PBE level lies in the zone of [0.26 eV, 1.63 eV] while the gaps in HSE06 level lies in [0.89 eV,2.44 eV]. For XA codoping configuration, the induced gaps in PBE level lies in the zone of [0.25 eV, 1.66 eV] while the gaps in HSE06 level lies in [0.29 eV,2.44 eV]. Generally, comparing to the PBE functional, the hybrid HSE06 functional introduces the bigger gap values to considered systems with the maximal amplitude of 242\% (AE case). Therefore, the use of HSE06 functional in this paper is essential to correct the the gaps from the PBE results. The maximum and minimum of gap occur in the AH (or HA) and EA cases, respectively. It's exciting that the gap for the most stable codoping site of the AB case is 1.08 eV, close to the well-known band gap of semiconductor Si ($\sim$1.14 eV), which may promise the potential applications of B/N codoping $\gamma$-graphyne in future.

%\section{Conclusion}
In summary, the geometric and electronic structures for B- or N- single doped and B/N-codoped in $\gamma$-graphyne are investigated from first-principles calculations. The pure GGA-type PBE functionals and the hybrid HSE06 functionals are adopted to determine the electronic properties of the considered doped systems. For single B or N doping, judging from the formation energy, the B doping is more easily to realize than the N doping in $\gamma$-graphyne. The B and N doping prefer to select different site of two inequivalent sites. Furthermore, the Fermi energy level move down or up after single B or N doping which gives us a chance to modulate the electronic properties of $\gamma$-graphyne. Therefore, we further investigate the geometric and electronic structures by B/N codoping into $\gamma$-graphyne. The results show the B-N pair is the most preferred codoping configuration. The Fermi energy level returns around the forbidden zone and the gaps exhibit an oscillated behavior with the increasing of distance between B and N atom. The oscillated band gap is caused by the collaborative effect between of the breaking of sublattice symmetry and energy level coupling in $\gamma$-graphyne.

%\section{Acknowledgments}
We gratefully acknowledge financial support from the Nation Science Foundation of China under grant No. 11304087 and 61401151, the Science and Technology Project of Hengyang City under grant No. 2013KJ33, and the Construct Program of the Key Discipline in Hunan province of China.

%%%%%%%%%%%%%%%%%%%%%%%%%%%%%%%%%%%%%%%%%%
\clearpage
\begin{table}
\caption{The distances between C and dopants ($d_{C-B}$ and $d_{C-N}$), the formation energies $E_{F}$, the vacuum energy level $E_{Vacuum}$, and the Fermi energy level $E_{Fermi}$ of B or N single doping. The $d_{C-B}$ or $d_{C-N}$ indicate the shortest bond when B or N atom bind with several C atoms. $W$ means work function and $W$=$E_{Vacuum}$-$E_{Fermi}$. $E_{Vacuum}$, $E_{Fermi}$ and $W$ for pristine $\gamma$-graphyne are 1.15, 1.89 and 5.17 eV, respectively }
\begin{ruledtabular}
\begin{tabular}{ccccccc}
 &\multicolumn{2}{c}{B doping}  &\multicolumn{2}{c}{N doping} \\
\cline{2-3} \cline{4-5}
                       & A site          &B site       &A site    &B site  \\
\hline
$d_{C-B(N)}$({\AA})    & 1.46      & 1.33   & 1.35     &1.19 \\
$E_{F}$(eV)         & 1.78            &2.78         & 3.51     & 2.47  \\
$E_{Vacuum}$(eV)  &1.12 &1.12 &1.16 &1.16\\
$E_{Fermi}$(eV)   &-4.22 &-4.20 &-3.43 &-3.21\\
$W$(eV)  &5.34&5.33&4.59&4.37 \\
\end{tabular}
\end{ruledtabular}
\label{table1}
\end{table}

\clearpage
\begin{table}
\caption{The distances between C and dopants ($d_{C-B}$ and $d_{C-N}$), the formation energies $E_{F}$, and the band gaps $E_{gap}$ for two configurations of B/N codoping. The $d_{C-B}$ or $d_{C-N}$ indicate the shortest bond when B or N atom bind with several C atoms. }
\begin{ruledtabular}
\begin{tabular}{cccccccc}
AX  &AB &AC &AD &AE &AF &AG &AH \\
\hline
$d_{C-B}$({\AA}) &1.51 &1.33 &1.53 &1.48 &1.47 &1.46 &1.47\\
$d_{C-N}$({\AA}) &1.17 &1.17 &1.34 &1.35 &1.18 &1.18 &1.35\\
$E_{F}$(eV)    &2.51 &3.25 &4.29 &3.56 &3.62 &3.43 &4.63 \\
$E_{gap}$(PBE)(eV)  &1.08 &0.54 &1.63 &0.26 &1.14 &0.57 &1.30 \\
$E_{gap}$(HSE)(eV)  &1.60 &0.96 &2.31 &0.89 &1.89 &1.01 &2.44 \\
\hline
XA  &BA &CA &DA &EA &FA &GA &HA \\
\hline
 $d_{C-B}$({\AA}) &1.33 &1.50 &1.49 &1.48 &1.33 &1.33 &1.47\\
$d_{C-N}$({\AA}) &1.41 &1.42 &1.40 &1.35 &1.39 &1.36 &1.35\\
$E_{F}$(eV)  &4.60 &5.51 &4.29 &4.63 &5.53 &5.89 &3.56   \\
$E_{gap}$(PBE)(eV) &1.18 &0.29 &1.63 &0.25 &1.29 &0.49 &1.66  \\
$E_{gap}$(HSE)(eV) &1.82 &0.29 &2.31 &0.66 &2.14 &0.49 &2.44   \\
\end{tabular}
\end{ruledtabular}
\label{table2}
\end{table}

\clearpage

\clearpage

\clearpage

\clearpage

%%%%%%%%%%%%%
\end{document}